\documentclass[aps,twocolumn,showpacs,superscriptaddress,pra]{revtex4-1}
\usepackage{bm,bbm}
\usepackage{amsmath,amssymb,amsfonts}
\usepackage{graphicx,times}

\begin{document}

\title{Comparing different non-Markovianity measures: \\ A case study}

\author{P. Haikka}
\email[]{pmehai@utu.fi}
\homepage[]{www.openq.fi}
\affiliation{Turku Center for Quantum Physics, Department of Physics and Astronomy, University of Turku, FIN-20014 Turku, Finland}
\author{J. D. Cresser}
\affiliation{CQST, Department of Physics and Astronomy, Macquarie University, Sydney NSW 2109, Australia}
\author{S. Maniscalco} 
\affiliation{Turku Center for Quantum Physics, Department of Physics and Astronomy, University of Turku, FIN-20014 Turku, Finland}
\email[]{smanis@utu.fi} \homepage[]{www.openq.fi}

\begin{abstract}
We consider two recently proposed measures of non-Markovianity applied to a particular quantum process describing the dynamics of a driven qubit in a structured reservoir. The motivation of this study is twofold: on one hand, we study the differences and analogies of the non-Markovianity measures and on the other hand, we investigate the effect of the driving force on the dissipative dynamics of the qubit. In particular we ask if the drive introduces new channels for energy and/or information transfer between the system and the environment, or amplifies existing ones. We show under which conditions the presence of the drive slows down the inevitable loss of quantum properties of the qubit.
\end{abstract}

\date{\today}

\maketitle

\section{Introduction}
The inevitable interaction of a quantum system with its environment leads to dissipation of energy and loss of quantum coherence \cite{breuer&petruccione, weiss}. Unfortunately these phenomena can fundamentally limit the potential of quantum technologies, whose power is based on quantum effects \cite{nilsen}. In suitable surroundings, however, the quantum system may temporarily regain some of the previously lost energy and/or information due to non-Markovian effects in the system dynamics. This is one of the reasons why the study of non-Markovian quantum dynamics has received an increasing amount of interest in the last several years \cite{list}. It is surprising, then, that the definition of non-Markovianity still remains elusive and, in some sense, controversial.\\
Markovianity is well defined for classical stochastic processes:  a stochastic process has the {\sl Markov property} if the probability distribution for the future states of the process depends only on the present state. Loosely speaking, dependence on past states should  then be a characteristic trait of non-Markovian processes. The mathematical definition of non-Markovianity does not, however, translate easily into the widely used language of density matrices and master equations in quantum physics. In the case of an open quantum system evolution described by a Lindblad master equation \cite{lindblad, gorini} one can find a stochastic descriptions for the system dynamics, e.g., with the Monte Carlo Wave Function method, where future dynamics of individual quantum trajectories only depend on their present states \cite{mcwf, mcwf2}. In this spirit some physicists have attributed non-Markovianity to the breakdown of the semi-group property \cite{eisert} or a generalization thereof, the divisibility property \cite{rivas}. However, finding a stochastic description that corresponds to more general master equations is not, in general, a straightforward task \cite{pilbr}.\\
With the aim of extending the concept of non-Markovianity to more general quantum dynamics, efforts have been made to clarify the very definition of non-Markovianity in the context of open quantum systems and to quantify the amount of non-Markovianity in a given quantum process. One viewpoint associates non-Markovianity to dynamical dependence on past evolution. This may mean either dependence of the evolution on all past states of the system, as in the memory kernel approach \cite{nakajima, zwanzig}, or a dependence of the asymptotic state of the system on its initial state \cite{pascazio}. Yet another definition equates non-Markovianity with a partial recovery of previously lost information \cite{info1, info2, fisher}. It is worth stressing that the definitions of non-Markovianity are generally not the same, nor are they mutually exclusive, and that they may agree perfectly about the non-Markovian character for some models and disagree about others \cite{comparison}.\\
In this paper we consider two recently proposed measures of non-Markovianity introduced by Rivas, Huelga and Plenio (RHP) \cite{rivas} and Breuer, Laine and Piilo (BLP) \cite{info1} and apply them to a quantum process describing the short and intermediate time-scale dynamics of a driven qubit in a structured reservoir \cite{physicascripta, PRA81}. The comparison allows, on one hand, to elucidate the  differences and analogies between the non-Markovianity measures in a physically interesting model, and on the other hand to gain new understanding on the effect of the driving force on dissipative dynamics. For the sake of concreteness in the rest of the paper we will refer to the case of a laser-driven atom. All the conclusions, however, are valid for any driven two-state system. Due to the presence of the drive, the commonly used secular approximation does not always hold, especially in the short non-Markovian time scales. The importance of the nonsecular terms has been recently demonstrated in the context of superconducting circuits in Ref. \cite{pekola1,pekola2}.

The article is organized as follows. In Sec. \ref{model} we introduce our model, namely the driven qubit, and the master equation describing its dynamics in a structured environment. In Sections \ref{RHP} and \ref{BLP} we introduce and study the RHP and BLP measures of non-Markovianity using this model. In Sec. \ref{comparison} we compare the non-Markovian properties of a driven qubit with those of a non-driven qubit and finally, in Sec. \ref{conclusion}, we summarize the results and present our conclusions. 

\section{The driven qubit \label{model}} 
We consider a qubit with energy separation $\omega_A$ ($\hbar=1$), driven by a laser of frequency $\omega_L$ almost resonant with the transition frequency of the qubit: $|\Delta|=|\omega_A-\omega_L|\ll\omega_A$. The qubit is embedded in a zero-T environment. The strengths of the interaction between the qubit and laser and the qubit and environment are quantified by the Rabi frequency $\Omega$ and the coupling constant $\alpha$, respectively. We assume that $\Omega\ll\omega_A$, a condition that is typically satisfied in quantum optical situations. When $\alpha$ is the smallest relevant frequency, that is, the qubit couples weakly to the reservoir, the dynamics of the qubit is given by the following local-in-time completely positive master equation, derived by one of us in Ref.  \cite{physicascripta}:
\begin{equation}
\label{meq}
\frac{d \rho(t)}{dt}=-i[H,\rho(t)]+\mathcal{D}_{S}(t)\rho(t)+\mathcal{D}_{NS}(t)\rho(t),
\end{equation}
where the unitary evolution of the qubit is generated by the system Hamiltonian
\begin{equation}
\label{ }
H=\frac{\omega}{2}\sigma_z,\quad \omega=\sqrt{\Delta^2+\Omega^2},
\end{equation}
and the dissipative terms have been divided into a secular contribution $\mathcal{D}_S$ and a nonsecular contribution $\mathcal{D}_{NS}$. Introducing the short-hand notation
\begin{equation}
\label{ }
\mathcal{L}\left[A(t),\gamma(t)\right]=\gamma(t)\left(A(t)\rho(t)A^\dagger (t)-\frac{1}{2}\{A^\dagger(t)A(t),\rho(t)\}\right)
\end{equation}
to describe a Lindblad-type decay channel with a jump operator $A(t)$ and a corresponding decay rate $\gamma(t)$, we can express the dissipative terms as:
\begin{eqnarray}
\mathcal{D}_{S}(t)\rho(t)&=&C_+^2\mathcal{L}[\sigma_-,\gamma_+(t)]+C_-^2\mathcal{L}[\sigma_+,\gamma_-(t)]\nonumber\\
&+&C_0^2\mathcal{L}[\sigma_z,\gamma_0(t)],\label{S}\\
\mathcal{D}_{NS}(t)\rho(t)&=&\Gamma_-(t)\Big\{C_-C_0\big[\sigma_+\rho(t)\sigma_z-\sigma_z\sigma_+\rho(t)\big] \nonumber \\
&+& C_+C_-\big[\sigma_+\rho(t)\sigma_+-\sigma_+\sigma_+\rho(t)\big]\Big\}\nonumber\\
&+& \Gamma_+(t)\Big\{C_+C_0\big[\sigma_-\rho(t)\sigma_z-\sigma_z\sigma_-\rho(t)\big] \nonumber \\
&+&C_+C_-\big[\sigma_-\rho(t)\sigma_--\sigma_-\sigma_-\rho(t)\big]\Big\}\nonumber\\
&+&\Gamma_0(t) \Big\{C_-C_0\big[\sigma_z\rho(t)\sigma_--\sigma_-\sigma_z\rho(t)\big] \nonumber \\
&+&C_+C_0\big[\sigma_z\rho(t)\sigma_+-\sigma_+\sigma_z\rho(t)\big]\Big\}+ H.c.\label{NS}
\end{eqnarray}
where $C_\pm=(\Delta\pm\omega)/2\omega$, $C_0=\Omega/2\omega$ and $H.c.$ denotes Hermitian conjugation. The master equation is given in the dressed state basis $|\psi_\pm\rangle=\pm\sqrt{C_{\pm}}|e\rangle+\sqrt{C_{\mp}}|g\rangle$, where $|e\rangle$ and $|g\rangle$ are the excited and ground states of the qubit. The decay rates in $\mathcal{D}_{NS}$ and $\mathcal{D}_{S}$ are connected via the relation
\begin{equation}
\label{ }
\Gamma_\xi(t)=\left[\frac{\gamma_\xi(t)}{2}-i\lambda_\xi(t)\right],
\end{equation}
where $\gamma_\xi, \lambda_\xi\in\mathbf{R}$ and $\xi\in\{-,0,+\}$. For the sake of concreteness we focus on a qubit embedded in a reservoir with a Lorentzian spectral density $J(\omega)=\alpha/(2\pi)\,\lambda^2/[\lambda^2+(\omega-\omega_0)^2]$, where $\omega_0$ is the center frequency and $\lambda$ the width of the Lorentzian. For this case the decay rates in Eqs. (\ref{S})-(\ref{NS})  take the form \cite{PRA81} 
\begin{eqnarray}
\label{ratel}
 \gamma_{\xi}(T)&=&\frac{\alpha^2 }{2(1+q_{\xi}^2)}\left(1-e^{-T}\cos q_{\xi}T+e^{-T} q_{\xi} \sin q_{\xi} T\right),\nonumber\\
 \lambda_{\xi}(T)&=&\frac{\alpha^2 }{1+q_{\xi}^2} \left(- q_{\xi} +e^{-T} q_{\xi} \cos q_{\xi} T+e^{-T}  \sin q_{\xi} T\right),\nonumber\\
\end{eqnarray}
where $T=\lambda t$ and $q_{\xi}=s-\xi p$, again with $\xi=\{-,0,+\}$. Different dynamical regimes are defined in terms of the parameters
\begin{equation}
s=\frac{\omega_0-\omega_L}{\lambda},\quad p=\frac{\tau_C}{\tau_S} =\frac{ \omega }{ \lambda}.
\end{equation}
The parameter $s$ quantifies the detuning between the qubit and the center frequency of the Lorentzian. The parameter $p$ expresses the relationship between the reservoir correlation time-scale $\tau_C=\lambda^{-1}$ and the typical system time-scale $\tau_S=\omega^{-1}$ and it determines the border between secular and nonsecular regimes. Note that here we refine the typical textbook notion of the secular approximation, which requires that the typical time-scale of the system is negligible in comparison with the relaxation time-scale and results in a coarse-graining of the relaxation dynamics. Since in this work we are interested in the short time-scale dynamics we assume a stronger condition, namely that the typical time-scale of the driven qubit is much smaller than the other time-scales and, in particular, the reservoir correlation time-scale $\tau_C$. We call the limit $\tau_S\ll\tau_C$ the strong secular limit.

\subsection{Strong secular limit}
When $p\gg1$  the secular approximation holds and  the nonsecular dissipation term $\mathcal{D}_{NS}$ can be neglected from Eq. (\ref{meq}). Consequently, in the language of Markovian \cite{mcwf, mcwf2} and non-Markovian \cite{nmqj, nmqj2} quantum jumps, the master equation comprises of three Lindblad-like terms describing phase flips and jumps between the eigenstates of the driven qubit, with the direction of the jump (regular/reversed) depending on the sign of the corresponding decay rate (positive/negative). We have shown in Ref. \cite{PRA81} that, in the secular regime, the decay rates $\gamma_\pm(t)$ always oscillate taking temporarily negative values, whereas $\gamma_0(t)$ is positive for small values of $s$ and has periods of negativity when $s\gtrsim3.6$.

\subsection{Nonsecular limit}
In the opposite regime, defined by $p\ll1$, we have to consider the full master equation including the nonsecular terms to obtain a proper description of the short time-scale dynamics. However, in this limit the decay rates take a simpler form. More precisely, $\gamma_\pm(t)\approx\gamma_0(t)\equiv\gamma(t)$ and $\lambda_\pm(t)\approx\lambda_0(t)\equiv\lambda(t)$ and again $\gamma(t)$ is positive for small values of $s$ and takes temporarily negative values when $s\gtrsim3.6$, while $\lambda(t)\leq0$ for all times $t$. With this approximation the master equation can be cast into a remarkably simple form with a single Lindblad-type decay channel:
\begin{equation}
\label{nsmeq}
\frac{d \rho(t)}{dt}=-i[H+H',\rho(t)]+\mathcal{L}[A, \gamma(t)],
\end{equation}
where the jump operator is
\begin{equation}
A=C_-\sigma_++C_+\sigma_-+C_0\sigma_z,
\end{equation}
and the additional term in the coherent evolution is
\begin{equation}
\label{ }
H'=\lambda(t)C_0(C_+-C_-)(\sigma_-+\sigma_+).
\end{equation}

\section{Semi-group and divisibility \label{RHP}}
Historically non-Markovianity has been closely associated to deviations from the Lindblad master equation. The completely positive trace preserving (CPTP) map $\Phi:\,\rho(0)\mapsto\rho(t)=\Phi(t)\rho(0)$ corresponding to a Lindblad master equation always satisfies the semi-group property $\Phi(t+s)=\Phi(t)\Phi(s)$. For example Wolf, Eisert, Cubitt and Cirac propose to call a map Markovian if it is a CPTP map satisfying the semi-group property \cite{eisert}. The master equation (\ref{meq}) for the driven qubit in any structured reservoir is not in the Lindblad form due to the time dependency of the decay rates and therefore the dynamical map is not an element of a single parameter semigroup and Markovian in this sense.\\
The measure for non-Markovianity proposed by Rivas, Huelga and Plenio is based on a more general property of maps than the semi-group property, namely divisibility \cite{rivas}. A CPTP map $\Phi(t+\tau,0)$ is divisible if it can be written as a decomposition of two CPTP maps $\Phi(t+\tau,t)$ and $\Phi(t,0)$:
\begin{equation} \label{concatenation}
\Phi(t+\tau,0)=\Phi(t+\tau,t)\Phi(t,0).
\end{equation}
When the dynamical map is homogenous in time, $\Phi(t+\tau,t)=\Phi(\tau,0)\equiv\Phi(\tau)$, Eq. (\ref{concatenation}) reduces to the semi-group property.\\
RHP define a map to be Markovian exactly when it is divisible. This amounts to requiring that $\Phi(t+\tau,t)$ is completely positive. It is shown in Ref. \cite{rivas} that the quantity
\begin{equation} \label{g}
g(t)=\lim_{\epsilon\rightarrow0}\frac{\|\Phi(t+\epsilon,t)\otimes \mathbf{I}\, |\phi\rangle \langle \phi |\|-1}{\epsilon}\geq0
\end{equation}
is strictly positive if and only if $\Phi$ is indivisible and hence $g(t)$ identifies non-Markovian processes, according to their definition. Here $|\phi\rangle=\frac{1}{\sqrt{d}}\sum_{i=0}^d|i\rangle| i\rangle$, where $d$ is the dimension of the quantum system; in our case $|\phi\rangle=(|11\rangle+|00\rangle)/\sqrt{2}$ is the maximally entangled Bell state. The quantity $g(t)$ gives rise to a measure for non-Markovianity defined as
\begin{equation} \label{divisibilitymeasure}
\mathcal{N}_{RHP}(\Phi)=\frac{\mathcal{I}}{\mathcal{I}+1},\quad\mathcal{I}=\int ds\, g(s),
\end{equation}
with the property that $\mathcal{N}_{RHP}\in[0,1]$. The lower and upper limits correspond to Markovian and maximally non-Markovian maps, respectively.\\
For a local-in-time master equation $d\rho(t)/dt=\mathcal{L}(t)\rho(t)$ the dynamical map, in the limit $\epsilon\rightarrow 0$, is formally given by $\Phi(t+\epsilon,t)=\exp[\mathcal{L}(t)\epsilon]$. Consequently, RHP propose that, in this case, one may  check the non-Markovianity of the dynamical process by means of a simplified quantity
\begin{equation}
\label{g2}
g(t)=\lim_{\epsilon\rightarrow0}\frac{\| \big\{\mathbf{I}+\epsilon[\mathcal{L}(t)\otimes\mathbf{I}]\big\} |\phi\rangle \langle \phi |\|-1}{\epsilon}.
\end{equation}

\subsection{Secular regime}
A direct calculation of Eq. (\ref{g2}) in the case of a driven qubit in the strong secular regime gives
\begin{eqnarray}
g(t)=\frac{C_+^2P[\gamma_+(t)] +C_-^2P[\gamma_-(t)] +2C_0^2P[\gamma_0(t)]}{2},
\end{eqnarray}
where we define an auxiliary function $P$: $P(x)=0$ when $x\geq0$ and $P(x)=-x$ when $x<0$. In general, whenever any of the decay rates is negative, divisibility is lost. In the Lorentzian case this is, indeed, the situation in the secular regime where $\gamma_\pm(t)<0$ always for some periods of time. Thus we prove that according to RHP the dynamics of the driven qubit in the secular regime and in the Lorentzian case, is always non-Markovian.

\subsection{Nonsecular regime}
In the nonsecular regime we find
\begin{eqnarray}
g(t)=\frac{(C_+^2+C_-^2+2C_0)P[\gamma(t)]}{2},
\end{eqnarray}
that is, the dynamics is indivisible and hence non-Markovian according to RHP whenever the decay rate $\gamma(t)$ takes temporarily negative values. Again the result is valid for any reservoir, in the weak coupling limit. Specifically, for a laser-driven qubit in a Lorentzian reservoir divisibility is broken whenever $s=(\omega_0-\omega_L)\lambda\gtrsim3.6$, that is, the laser and the atom are sufficiently detuned from the center of the Lorentzian, as discussed in Sec. \ref{model}. 

\section{Information backflow \label{BLP}}
As a second measure of non-Markovianity we consider the one proposed by Breuer, Laine and Piilo in Ref. \cite{info1}.  The BLP measure aims at identifying non-Markovian dynamics with certain dynamical physical features of the system-reservoir interaction, rather than with the mathematical properties of the dynamical map of the open system. In particular, BLP define non-Markovian dynamics as a time evolution for the open system characterized by a temporary flow of information from the environment back into the system. This regain of information manifests itself as an increase in the distinguishability of pairs of evolving quantum states. Hence, the dynamical map $\Phi$ is non-Markovian according to BLP if there exists a pair of initial states $\rho_1(0)$ and $\rho_2(0)$ such that for some time $t$ the distinguishability of $\rho_1(t)$ and $\rho_2(t)$ increases, that is,
\begin{equation} \label{sigma}
\sigma(\rho_1,\rho_2,t)=\frac{d}{dt}D[\rho_1(t),\rho_2(t)]>0,
\end{equation}
where $D(\rho_1,\rho_2)=\frac{1}{2}\|\rho_1-\rho_2\|$ is the distinguishability of $\rho_1$ and $\rho_2$ with $\|A\|=\sqrt{AA^\dagger}$ being the trace distance, and $\rho_{1/2}(t)=\Phi(t,0)\rho_{1/2}(0)$.
When the distinguishability of two states increases information flows from the environment back to the system and {\sl vice versa}. With this definition one can quantify the amount of non-Markovianity in the quantum process as follows
\begin{equation} \label{measure1}
\mathcal{N}_{BLP}(\Phi)=\max_{\rho_{1,2}(0)}\int_{\sigma>0}dt\,\sigma(\rho_1,\rho_2,t),
\end{equation}
that is, $\mathcal{N}_{BLP}$ gives the maximum amount of information that can flow back to the system for the given process.\\
All divisible maps continuously reduce the distinguishability of quantum states, that is, for a divisible $\Phi$ we get $\mathcal{N}_{BLP}=0$. Hence the connection between the BLP and RHP measures is the following: if a map is Markovian according to RHP, it is Markovian also according to BLT. The converse is  in general not true.

\subsection{Secular regime}
A straightforward application of Eq. (\ref{sigma}) to the solution of the master equation (\ref{meq}) in the secular regime gives
\begin{equation}\label{secsig}
\sigma(t)=-\frac{e^{-2\Lambda(t)}\Lambda'(t)(\delta x^2+\delta y^2)+e^{-2\Gamma(t)}\Gamma'(t)\delta z^2}{2\sqrt{e^{-2\Lambda(t)}(\delta x^2+\delta y^2)+e^{-2\Gamma(t)}\delta z^2}},
\end{equation}
where $\delta x=x_1(0)-x_2(0)$, $\delta y=y_1(0)-y_2(0)$, and $\delta z=z_1(0)-z_2(0)$ are the differences in the $x$-, $y$- and $z$- components of the Bloch vector representations of $\rho_1(0)$ and $\rho_2(0)$, respectively, and
\begin{eqnarray}\label{Gamma}
\Gamma(t)&=&\frac{1}{2}\int_0^t ds \left[C_+^2\gamma_+(s)+C_-^2\gamma_{-}(s)+4C_0^2\gamma_0(s)\right],\nonumber\\
\Lambda(t)&=&\int_0^t ds \left[ C_+^2\gamma_+(s)+C_-^2\gamma_{-}(s)\right]. \label{Lambda}
\end{eqnarray}
The expression in Eq. (\ref{secsig}) is valid for a generic structured reservoir and the non-Markovian properties of the driven qubit strongly depend on the choice of the environment. We now focus on the Lorentzian case. Let us choose a pair of initial states such that $\delta z=0$. Then the sign of $\sigma$ only depends on $\Lambda'(t)=C_+^2\gamma_+(t)+C_-^2\gamma_-(t)$. Recall from Sec. \ref{model} that in the secular regime $\gamma_\pm(t)$ have negative periods for all possible parameter values. Then $\sigma(t)>0$ always for some periods of time. Therefore, in the secular regime the process describing the dynamic of a driven qubit in a Lorentzian reservoir is always non-Markovian according to BLP.\\
It is worth stressing that although the BLP and RHP measures both agree that the dynamics of the driven qubit in a Lorentzian reservoir is non-Markovian, the result need not hold when the driven qubit is embedded in some other structured reservoir. Indeed, negativity of one of the decay rates does not guarantee that $\sigma(t)>0$, although it immediately breaks the divisibility.  It is therefore natural to ask if we can find some general parameter regions when the dynamics of the driven qubit is Markovian according to BLP, but non-Markovian according to RHP. It turns out that the question is non-trivial from many points of view. For one, showing that the reduced dynamics of an open system is non-Markovian according to BLP is simple in comparison to showing that the dynamics is Markovian, since the former case amounts to finding one pair of initial states whose distinguishability increases for some time interval, while the latter case requires that one probes all possible pairs of states and shows that distinguishability can only decrease. Such optimization procedure is possible in practice only numerically, making it hard to isolate parameter regimes for different dynamical behavior. Moreover, the BLP non-Markovianity quantifier depends strongly on the spectral density of the environment via linear combinations of the three decay rates and their integrals, so identifying the parameters for which the two measures deviate may not even be possible unless one has a specific environmental spectral density in mind. Moreover, even specifying the spectral density, one needs an extensive numerical study involving several variables. So far we have been unable to find instances when the two measures disagree and the question remains open.

\subsection{Nonsecular regime}
In the nonsecular regime the master equation (\ref{meq}) is not, in general, solvable analytically and we can evaluate the quantity $\sigma$ only numerically. We can, however, consider the special case of resonance between the atom and the laser. In this case the master equation is solvable and we find
\begin{equation}\label{resonantsigma}
\sigma(t)=-\frac{\gamma(t)e^{-2\int_0^tds\gamma(s)}[(\delta x^2-\delta z^2)+2\delta y^2]}{\sqrt{2}\sqrt{(\delta z^2+\delta x^2)+e^{-2\int_0^tds\gamma(s)}[(\delta x^2-\delta z^2)+2\delta y^2]}},
\end{equation}
Thus in the resonant nonsecular case the process is non-Markovian according to BLP if and only if $\gamma(t)<0$ for some time $t$. Numerical studies of the off-resonant case $\Delta\neq0$ indicate that this result holds more generally, that is, in the nonsecular regime $\mathcal{N}_{BLP}(t)=0$ if and only if $\gamma(t)\leq0$.\\

\section{Laser induced non-Markovianity \label{comparison}}
In this section we compare the non-Markovian character of the driven dissipative qubit and the  analytically solvable model of an unperturbed dissipative qubit when both are embedded in a Lorentzian reservoir \cite{breuer&petruccione,sab1}. The master equation for the qubit in a Lorentzian reservoir, in the special case when the qubit is in resonance with the Lorentzian, $\omega_A=\omega_0$, is
\begin{eqnarray}
\frac{d \rho(t)}{dt}=\gamma(t)\left[\sigma_-\rho(t)\sigma_+-\frac{1}{2}\{\sigma_+\sigma_-,\rho(t)\}\right],
\end{eqnarray}
where, in contrast to the master equation of Eq. (\ref{meq}), the operators are in the bare state basis $\{|e\rangle,\,|g\rangle\}$. The time-dependent decay rate is
\begin{equation}
\gamma(t)=\frac{2 \alpha \lambda \sinh(dt/2)}{d \cosh(dt/2)+\lambda \sinh(dt/2)},
\end{equation}
where $d=\sqrt{\lambda^2-2\alpha \lambda}$. Applying this model to Eqs. (\ref{sigma}) and (\ref{g2}) and introducing quantity $\Gamma(t)=\int_0^tds\gamma(s)$we get
\begin{eqnarray}
\sigma(t)&=&-\frac{\gamma(t)\left[2e^{-2\Gamma(t)}\delta z^2+e^{-\Gamma(t)}(\delta x^2+\delta y^2)\right]}{4\sqrt{e^{-2\Gamma(t)}\delta z^2+e^{-\Gamma(t)}(\delta x^2+\delta y^2)}},\label{eq:23}\\
g(t)&=&\frac{1}{2}P[\gamma(t)]\label{eq:24}.
\end{eqnarray}
Equations (\ref{eq:23})-(\ref{eq:24}) show that, in absence of the driving laser, the dynamics is non-Markovian for both the RHP and the BLP measures if and only if $\gamma(t)<0$ for some time $t\in\mathbf{R}_+$. This happens exactly when $\lambda<2\alpha$, that is, when the qubit couples strongly to the environment.\\
Consider now a situation where the qubit couples weakly to the Lorentzian reservoir and the transition frequency of the qubit is resonant with the center of the Lorentzian. The analysis above shows that, without the laser drive, the dynamics of the qubit is Markovian according to both RHP and BLP. If, instead, we drive the qubit with a laser, the results obtained in Secs. \ref{BLP} and \ref{RHP} show that we can induce non-Markovianity in the system dynamics when the laser parameters are suitably chosen. More precisely, the laser should couple to the qubit in such a way that condition $p\gg1$ holds, that is, we are in the secular regime. The secular regime can be achieved with a large enough Rabi frequency $\Omega$ and/or when the laser -and consequently the atom- are sufficiently detuned from the center of the Lorentzian.  When this happens the laser interferes with the dynamics of the dissipative qubit so much that it can temporarily reverse the flow of information from the qubit to the environment. 

\section{Discussion and conclusions \label{conclusion}}  
We have studied the non-Markovian character of a driven qubit in a structured reservoir in different dynamical regimes, using two different non-Markovianity measures, and we have analyzed the effect of the driving laser on the non-Markovian properties of the dissipative qubit.\\
In the secular regime both measures confirm that the dynamics of the driven qubit is always non-Markovian when it is embedded in a Lorentzian reservoir. However, in general this need not be the case and deviations between the two measures could be discovered when the driven qubit is embedded in another reservoir. So far we have been unable to find a spectral density for which the two measures disagree. This might be an indication that there are no realistic physical situation where this happens, even if the mathematical definitions of the quantifiers suggest this. It remains an interesting open question whether the two non-Markovianity measures always agree in the case of a driven qubit embedded in a structured reservoir; indeed deviations between the two measures are neither characterized nor well understood to date.\\
Instead in the nonsecular regime the non-Markovianity measures agree perfectly. In this regime the appearance of non-Markovian features is strongly dependent on the way the laser is coupled to the qubit. Neither the presence of the structured reservoir nor the presence of the laser-drive is sufficient to guarantee non-Markovianity in the system dynamics in the nonsecular regime. In Sec. \ref{comparison} we investigated the origins of these non-Markovian effects by comparing driven and non-driven qubits in a Lorentzian reservoir; it was shown that non-Markovian effects are not possible without a strong laser-drive and/or detuning between the driving laser and the center frequency of the Lorentzian distribution. This discovery has a clear physical interpretation; only a strong enough laser drive can induce a feedback of information from the environment into the system. In fact, non-Markovianity in the nonsecular regime occurs in the far off-resonant case. Far off-resonance dynamics is usually associated to the presence of virtual processes which may be at the origin, in this case, of the flow-back of information.

\begin{acknowledgments}
This work was supported by the Emil Aaltonen foundation, the Finnish Cultural foundation, and by the Turku Collegium of Science and Medicine (S.M.). P.H. is grateful to \'Angel Rivas for many useful comments and discussions. We also acknowledge stimulating discussions with J. Piilo.
\end{acknowledgments}

\end{document}